      [1999/12/01 v1.4c Il Nuovo Cimento]
\documentclass{cimento}

\usepackage{graphicx}  

\title{The four-site Higgsless model at the LHC}
\author{E.~Accomando\from{ins:infn}\ETC,
S.~De~Curtis\from{ins:infn},
D.~Dominici\from{ins:univ},
L.~Fedeli\from{ins:univ}}
\instlist{\inst{ins:infn} I.N.F.N., Sezione di Firenze, I-50019 Sesto F., Italy
  \inst{ins:univ} Dipartimento di Fisica, Universit\`a di Firenze, I-50019 Sesto F., Italy}
\PACSes{\PACSit{11.15.Ex}\PACSit{12.60.Cn}}
\begin{document}

\maketitle

\begin{abstract}
We consider the four-site Higgsless model, 
which 
predicts the existence of four charged and two neutral extra gauge bosons, 
$W_{1,2}^\pm$ and $Z_{1,2}$. In contrast to other Higgsless models,
characterized by fermiophobic extra gauge bosons, here sizeable fermion-boson 
couplings are allowed by the electroweak precision data. We thus analyse the 
prospects of detecting the new predicted particles in the mostly favored 
Drell-Yan channel at the LHC.  
\end{abstract}

\section{Introduction}

In recent years, a renewed attention has been focused on Higgsless models 
\cite{higgsless}. They emerge naturally from local gauge theories in five 
dimensions, and their major outcome is delaying the unitarity violation of 
vector-boson scattering (VBS) amplitudes to high energies by the exchange of 
Kaluza-Klein excitations \cite{unitarity}. Their common drawback is to 
reconcile unitarity with the ElectroWeak Precision Test (EWPT) bounds.

Within this framework, and in the attempt to solve this dichotomy, many models 
have been proposed. We concentrate on the so called deconstructed 
theories \cite{deconstructed} which come out from the discretization of the 
fifth dimension to a 
lattice, and are described by chiral lagrangians with a number of gauge-group 
replicas equal to the number of lattice sites. The simplest version of 
this class of models is related to the old BESS model \cite{bess}, a lattice 
with only three sites 
and $SU(2)_L\times SU(2)\times U(1)_Y$ gauge symmetry (for it, sometimes 
called three-site Higgsless model). In view of the LHC experiments, its 
phenomenological consequences have recently received a renewed attention. The 
investigated extra-gauge-boson production channels require however high 
luminosity to be detected. In order to reconcile unitarity and EWPT-bounds, 
this minimal 
version predicts indeed the new triplet of vector bosons to be almost 
fermiophobic. Hence, only di-boson channels, vector boson fusion and 
triple gauge boson production processes have been analysed 
\cite{Higgsless-pheno}. 

We extend the minimal model by inserting an additional lattice site. This new 
four-site Higgsless model \cite{foursite,pap}, based on the 
$SU(2)_L\times SU(2)_1\times SU(2)_2\times U(1)_Y$ gauge symmetry, predicts 
four charged and two neutral extra gauge bosons, $W_{1,2}^\pm$ and $Z_{1,2}$, 
and satisfies the EWPT bounds without forcing the new resonances to be 
fermiophobic. As a consequence, they could be detected at the LHC in the 
promising Drell-Yan channel.

\section{Unitarity and EWPT bounds}

\begin{figure}
\centering
\vskip -3.8cm
\includegraphics[height=25.cm]{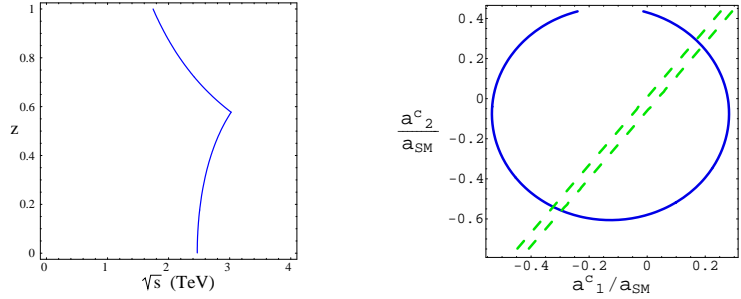} 
\vskip -17.cm    
\caption{Left) Unitarity bound vs energy scale for different $z=M_1/M_2$ 
values. The perturbative region is on the left of the curve. 
Right) 95$\%$ C.L. bounds on the charged fermion-boson couplings normalized to 
the SM values, from $\epsilon_1$ (solid) and $\epsilon_3$ (dashed), and 
$M_1=1$ TeV, $M_2=1.3$ TeV.}
\label{fig:1}
\end{figure}

\noindent
The four-site Higgsless model is described by four parameters: 
$b_1, b_2, M_1, M_2$. These are related to the two direct couplings between 
$V_{1,2}$-bosons ($V=W,Z$) and SM fermions and the two $V_{1,2}$ masses 
(charged and neutral gauge bosons are degenerate), respectively. The 
introduction of $b_{1,2}$ represents
the novelty of the proposed model. It allows to reconcile unitarity
and EWPT bounds, leaving a calculable and not fine-tuned parameter space.

The energy range, where the perturbative regime is still valid 
is plotted in Fig.1L for different values of $z=M_1/M_2$. Owing to the
exchange of the extra gauge bosons, the unitarity violation can be delayed up 
to an energy scale of about $\sqrt{s}\simeq 3$ TeV. 
Hence, the mass spectrum of the new particles is constrained to be within a 
few TeV. 

In the literature, the only way to combine the need of relatively low mass 
extra gauge bosons with EWPT was to impose the new particles to be 
fermiophobic. In the four-site Higgsless model, this strong assumption is not 
necessary anymore. In Fig.1R, we plot the bounds on the charged fermion-boson 
couplings from the EWPT expressed in terms of the $\epsilon_{1,3}$ parameters 
. The outcome is
that $\epsilon_3$ constraints the relation between the two couplings, while
$\epsilon_1$ limits their magnitude. Neutral couplings share a similar
behaviour. The allowed fermion-boson couplings are thus bounded, but still
sizeable. We can then explore their phenomenological consequences in Drell-Yan
production at the LHC.

\section{$W^\pm_{1,2}$ and $Z_{1,2}$ Drell-Yan 
production at the LHC} 
\begin{figure}
\centering
\vskip -1.8cm
\includegraphics[height=20cm]{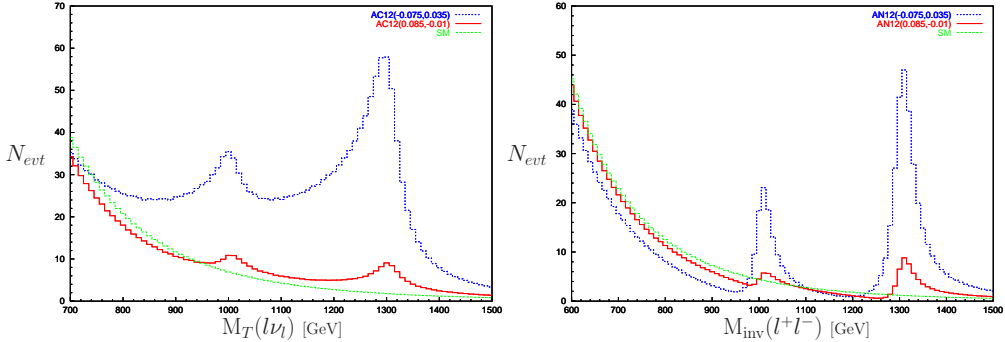}
\vskip -13.2cm
\caption{Left) Total number of events as a function of the transverse mass for 
the process $pp\rightarrow l\nu_l$. We sum over $e,\mu$ and charged conjugate 
processes. Right) Total number of events versus the dilepton invariant 
mass for the process $pp\rightarrow l^+l^-$. We sum over $e,\mu$.
In both figures, we take $M_1=1$ TeV, $M_2=1.3$ TeV and L=10 fb$^{-1}$. 
Standard cuts are applied, legends as in the text.} 
\label{fig:2}
\end{figure}

In this section, we discuss the prospects of detecting the six extra gauge bosons at the LHC. Our numerical setup is here summarized:
$M_Z=91.187$ GeV, $\alpha (M_Z)=1/128.8$, $G_\mu =1.1664\cdot 10^{-5} GeV^{-2}$. 
We adopt the fixed-width scheme and apply standard acceptance cuts: 
$P_t(l)>20$ GeV, $P_t^{miss}(l)>20$ GeV, $\eta (l)<2.5$. The CTEQ6L is used.
As an example of the four-site Higgsless model prediction, we choose
two sets of free parameters: $(b_1, b_2)=(-0.075,0.035)$ and 
$(b_1, b_2)=(0.085,-0.01)$ at fixed values $M_1=1$ TeV, $M_2=1.3$ TeV. We 
compute the full Drell-Yan process, considering signal and SM-background, at 
EW and QCD leading order. A luminosity L=10 fb$^{-1}$ is assumed.

In Fig. 2, we analyse both charged and neutral Drell-Yan channels. 
From top to bottom, the three curves represent the first setup, the latter and
the SM prediction. In Fig. 2L, we plot the number of events in a 10-GeV bin as 
a function of the leptonic transvers mass, $M_t(l\nu_l)$, for the process 
$pp\rightarrow l\nu_l$. Fig. 2R shows instead the number of events in a 10-GeV 
bin as a function of the leptonic invariant mass, $M_{inv}(l^+l^-)$, for the 
process $pp\rightarrow l^+l^-$.
The number of signal (total) events in the range 
$|M_{inv}(l^+l^-)-M_{1,2}|<\Gamma_{1,2}$ for the neutral channel is given by
$N_{evt}(Z_1)=83(96)$ and $N_{evt}(Z_2)=242(254)$ for the first setup.
The results show that, while the second setup would need high luminosity, in
the first one the new gauge bosons could be discovered already at the LHC 
start-up, with a minimum luminosity L=1fb$^{-1}$.

\section{Conclusions}
The four-site Higgsless model is calculable and not fined-tuned. It predicts
the existence of six extra gauge bosons, $W^\pm_{1,2}$ and $Z_{1,2}$, which
can have sizeable couplings to fermions. They could thus be seen in the 
favoured Drell-Yan channel at the LHC, even during its early-stage run at low 
luminosity. A detailed analysis will be given in Ref.\cite{pap}.


\end{document}